\documentclass{llncs}

\usepackage{subfig}
\usepackage{amsmath}
\usepackage{graphicx}

\begin{document}

\title{Accelerating System Log Processing by Semi-supervised Learning: A Technical Report}

\author{Guofu Li\inst{1},
	Pengjia Zhu\inst{2}\thanks{Work done while in AI Lab of Accenture China. }, 
	Zhiyi Chen\inst{1},\\
	}
	

\institute{
University of Shanghai for Science and Technology, \\College of Communication and Art Design, Shanghai, China\\
\email{li.guofu.l@gmail.com}, \email{iamedithchen@gmail.com}\\
\and
State Street Corporation, Hangzhou, China \\
\email{zhupengjia@gmail.com} \\
}

\maketitle

\begin{abstract}

	There is an increasing need for more automated system-log analysis tools for large scale online system in a timely manner. 
	However, conventional way to monitor and classify the log output based on keyword list does not scale well for complex system in which codes contributed by a large group of developers, with diverse ways of encoding the error messages, often with misleading pre-set labels. 
	In this paper, we propose that the design of a large scale online log analysis should follow the ``Least Prior Knowledge Principle'', in which unsupervised or semi-supervised solution with the minimal prior knowledge of the log should be encoded directly. 
	Thereby, we report our experience in designing a two-stage machine learning based method, in which the system logs are regarded as the output of a {\em quasi-natural language}, pre-filtered by a perplexity score threshold, and then undergo a fine-grained classification procedure. 
	Tests on empirical data show that our method has obvious advantage regarding to the processing speed and classification accuracy. 

~\\


\end{abstract}


\section{Introduction}

	Large scale online system outputs tons of log data every day, containing various types of the sensitive and valuable information that requires the attention from the system administrators. In many of the scenarios, analyzing the log data by human labour is never an option, either due to its vast volume, or its real-time reaction that is needed by the business. Therefore, creating a well-designed, automated process to monitor and classify this large volume of log data is becoming a growing need with the arrival of age of big-data. However, building such a system are facing several key challenges. 
	
	A simple and intuitive way to handle these system logging automatically is by setting a list of keywords, based on which either online monitoring or off-line analysis can be easily implemented. However, the logging messages that set by the system developers are often similar to a type of natural language used in a special-domain, with semantics often hard to be recovered by a simple set of word-based filter rules, when the word-choice made by the system exceptions of different types are often with large overlapping, full of the names of the sub-modules or functions, making it even harder to separate by keywords. 
	Though many of the built-in logging modules provide the system developers with extra functionalities to include meta-information (e.g., a set of tags, warning levels, or the logging output time) other than the basic disk-writing, a considerable portion of the coding practitioners do not have full consensus of the exact semantic of these labels, often follow their own way, and can even break the self-consistency in different part of the program. In practice, we observe a large amount of entries with label error which are actually normal, when actual error messages may be labeled as normal. 
	
	Another key challenge that requires careful examination by the designer of the the log analysis tools is that the logging output volume commonly grows with system size and complexity. 
	For a large commercial system, it is uncommon to see several hundreds Tera-bytes of its historical logging data waiting to be processed, and Gigs of new logging data are being produced every day. To ensure that the newly generated logging data do not stacked and accumulated on the disk, the analysis algorithm has to be at least as fast as the generation speed, but with more limited computation power. 
	This restriction blocks the log analyzer's developer from the choice of many time-consuming machine-learning algorithms, when still facing a complexity of the blurry semantics within the vast amount of data.
	
	In addition to the previous two challenges, we emphasize that the develop team of the system logging analyzer is usually different from the one who build the target system\footnote{We use the term ``target system'' to refer to the system which produce the logging data to be analyzed.} in the first place. Consequently, it is difficult, or sometimes impossible, for the algorithm designer of the logging analyzer to have full-knowledge of the conditions and implications of the logging messages. Besides, when a team is dedicated to the development of the logging analyzer, its goal is often to make the algorithm as generically applicable as possible. Based on this consideration, we propose a ``least prior knowledge'' designing principal of such tools, which aims to leverage the large volume of the unlabeled logging data, with the minimum supervision.\footnote{It may be confusing to use the term ``unlabeled'', as the log messages commonly carry labels when the are firstly generated, which may highly obscure and unrelated to their actual meaning.} 

	This report shares our working experience on a log-analysis system, design for two business modules of a Chinese telecom company branch. We choose a language model to first pre-filter a large amount of trivial log messages for acceleration, then a topic model based on the LDA (Latent Dirichlet Allocation) is used to before we use feedforward network for supervised classification. Though these methods may seem dated from today's perspective of natural language processing, they serve the purpose well and help to achieve a good balance between the speed and accuracy. 

\section{Backgrounds}

	Many of the conventional log analysis tools are built from a simple keyword list. For one reason, the system usually has a fixed template to output its runtime exception. And for another, each of the system developer usually has a style of output a certain type of error or message with a relatively fixed pattern. Therefore, for a small scale system, keywords do reflect the natural of the system logging, and implementing such an analysis tool via keywords has a good balance between the cost and effectiveness. However, as the system grows larger and the service it provides grows complex, this approach gradually becomes infeasible in practice. 

	\subsection{Related Works}

	Researches on analyzing system logs automatically evolves along with the evolvement of the target systems themselves, including their types, volume size, etc. For instance, 
Silverstein et al. studies the task of query log analysis at web scale \cite{silverstein1999analysis}. They analysis of an AltaVista Search Engine query log to study the interaction of terms within queries, aiming to find correlations between the searching items. 
	Splunk\footnote{https://www.splunk.com/en\_us/homepage.html} is one of the most widely used tool for log analysis. It provides several out-of-box services that are key to the system administrators with several built-in machine learning algorithms, and has been integrated inside the Google's cloud platform. However, Splunk only provides a command line wrapper of a few general-purpose machine learning algorithms that have not been tailored for log analysis tasks yet. 

	The runtime monitoring system proposed by Xu et al. \cite{xu2009detecting}, is another representative example to apply machine learning techniques on log data mining, which involves four sub-modules of {\em log parsing}, {\em feature creation}, {\em anomaly detection} and {\em visualization}. 
	The Sher-Log tool proposed by Yuan et al. \cite{yuan2010sherlog} shares a similar design principle of our method. It analyzes source code by leveraging information provided by run-time logs, and predicate the events that would happen in the production run without re-executing the program or prior-knowledge on the log's semantics. 
	The {\em Beehive} project by Yen et al. \cite{yen2013beehive} targets at the network security issue, and aims to extract useful knowledge by mining the dirty log data. Their experiments are conducted on the network packages and regards the meta-information of different layers of the network protocols as the mean features for classification. 
	Kobayashi et al. \cite{kobayashi2014towards} suggest that processing the log data can be regarded as a special type of natural language processing problem, such that many of the NLP methods can be used for analyzing the data and extracting a repository of templates. 

	We share a similar intuition that regards log messages as quasi-natural language, and make use of the language model \cite{charniak1996statistical} to perform pre-filtering. 
	Language model has been applied to various types of natural language related tasks. For instance, Ponte et al. use a language model to improve the information retrieval system \cite{ponte1998language}. 
	Similar to our idea, Salvetti et al. \cite{salvetti2006weblog} applies the language model to another quasi natural language scenario, web URLs, for segmentation and analysis in a fast weblog classification task. 

	Du et al. \cite{du2017deeplog} are among the first who apply deep learning methods on log data analysis, and proposed a system called {\em DeepLog}, which also regards the log data as a sequence of natural language output, and relies on a LSTM-based network architecture to find new patterns in the logs. 
	Their LSTM-based approach is also motivated by the language model, similar to a part of our work, as the optimization objective is also to predicte the next token based on the previously seen context. However, their work focuses on a different set of tasks, which are anomaly detection and root cause analysis.

	\subsection{Task Description}

	The logging messages written by an online service system is a rich repository of the event history of the system, as well as an indicator of the current status of service. They provide means more than the debug information for the system develop team, but also the source of intelligence for the system maintenance and audit. 
	Therefore, various tasks can be defined as the goal of the log analyzer. In this paper, we consider the goal to be a multi-way classification, which is a common setting. More specifically, we consider three classes of the messages: 
	\begin{enumerate}
	\item {\em information message}, 
	\item {\em operation error message}, and 
	\item {\em system error message}.
	\end{enumerate}
	Note that the actual class label that each message should be assigned to is often different from what it received from the target system, though the labeling schema may seem similar to each other. For example, a ``{\tt Connection Timeout Exception}'' may be set as a {system error} by the target system, but is actually harmless to the entire work flow of the system. It may be just a bad coding practice in the target system's code for handling the exception. 
	
	It is common for an online service system to run in $24 \times 7$ schedule, while the logging keeps constantly growing. In our specific scenario, the logging data that output from only two of the sub-modules that we experiment with are about 5GB and 90GB each day, when the volume of the logging data from all the sub-modules grows up-to 3TB each day.  
	Thus, it imposes a type of constraint that the average processing speed for the analysis tool must be faster than the average speed of the growth of the log data. Therefore, our approach splits the whole process into two consecutive stages: 
	\begin{enumerate}
		\item first, we use a language model to quickly filter out a large portion of the low-value messages;
		\item then, we use a three-way classifier based on LDA topic modeling and feedforward neural networks. 
	\end{enumerate}

\section{Language Model Based Pre-Filter}

	The goal of the first stage of our method is to perform a fast pre-filtering on the large volume of logging data, similar to the ``roll-out policy'' in a Monte-Carlo simulation in reinforcement learning algorithm. 

	\subsection{Fitting the Language Model}

	The term ``language model'' in the context of common NLP (Natural Language Processing) refers to a way to find the joint probability of a word (token) sequence $P(w_{1:T})$, by breaking the join probability to the product of a chain of conditional probabilities:
	$$ P(w_{1:T}) = \prod_{t=1}^{T}P(w_t | w_{1:t-1}) \text{,}$$ 
	in which $w_{1:T}$ denotes the word-sequence from position $1$ to $T$. 
	In many of the circumstances, we refer to the sequence behind the condition bar as the {\em context} that the probability is conditioned on. With an n-order Markovian assumption, we may have
	$$ P(w_t | w_{1:t-1}) \simeq P(w_t | w_{t-n:t-1}) \text{,}$$
	such that 
	$$ P(w_{1:T}) \simeq \prod_{t=1}^{T}P(w_t | w_{t-n:t-1}) $$
	
	Recent researches in the NLP field have witnessed the shift of the language model from the {\em statistic} language model to the {\em neural} language model, in which distributed representation is used to represent the semantic of each word or token. 
	Comparatively, traditional statistic language model is a non-parametric model that trivially estimates the probabilities based on the word counts in the training corpus, while the neural language model creates an embedding layer (thus parametric) on which semantic of the context is modeled. 
	Although empirical studies shows that the neural language model has advantages in smoothing and generalization, calculating the conditional probability by looking up a count-table is much faster than running a forward pass in a deep neural network (usually some variations of the recurrent neural network is adopted for these tasks). 
	Therefore, in our experiment, we choose to follow a traditional 3-gram language model to perform a fast pre-filtering. Comparing to the keyword-based filter, language model has the capability to take the word-order into consideration, making it ideal to handle the log messages whose meaning is more obvious by recognizing its template rather than its wording. 
	
	Though language model itself can be trained to make text classification directly, our approach however considers the perplexity derived from it to be an indicator of how common a certain type that the message in question is seen in the training corpus. It can be counter-intuitive that the most notable log (error) message that the system administrator should pay attention to, is the message whose type is rarely seen previously. However, considering a system that has been running stably for a sufficient long time, we should have the confidence that it behaves as expected in most cases. Therefore, regardless of whether the message was labeled as ``error'' by the target system, we care more about the ``unusual'' events, while the ``commonly-seen'' messages are trivial in natural. 
	
	\subsection{Filtering by Perplexity}
	
	Perplexity (or $ppx$) is a score implied by the language model, meaning the averaged joint probability of the token sequence when trained on corpus. Formally speaking, we may calculate the $ppx$ score by
	$$ ppx = P(w_{1:T}) ^ {\frac{1}{T}} \text{.}$$ 
One way to interpret the $ppx$ as the ``average branching factor'' of a long sequence given by a language model. Thus, a less ambiguous sequence prefix always implies a more steep distribution of the next token. 
	Since the impact of the length of the sequence is removed from the $ppx$ score, with a fixed model and training corpus, the relative $ppx$ difference is manly determined by how familiar certain sequence of tokens are observed in the training data, while the absolute values of the $ppx$ is less interesting to our task. 
	Often, we use a base-2 log-scaled ppx for simplicity:
	$$ \log(ppx) = \frac{1}{T} \log P(w_{1:T}) \simeq \frac{1}{T} \sum_{t=1}^{T} \log P(w_t | w_{t-n:t-1}) $$
	
\begin{figure}[t]
	\centering
    	\subfloat[The {\em scrmapp} submodule. ]{
    	\includegraphics[width=0.45\textwidth]{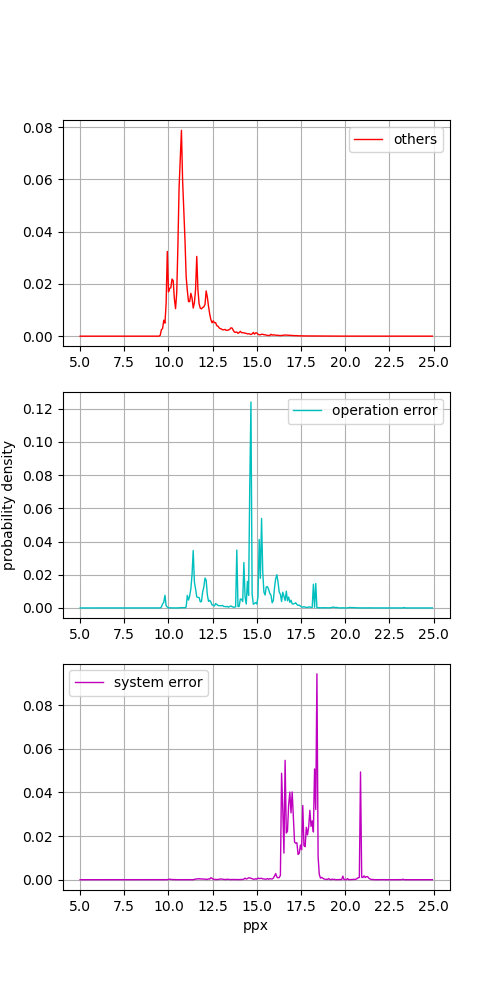}
    	\label{fig:scrmapp}
		}
		\hfil
		\subfloat[The {\em scrmweb} submodule. ]{
		\includegraphics[width=0.45\textwidth]{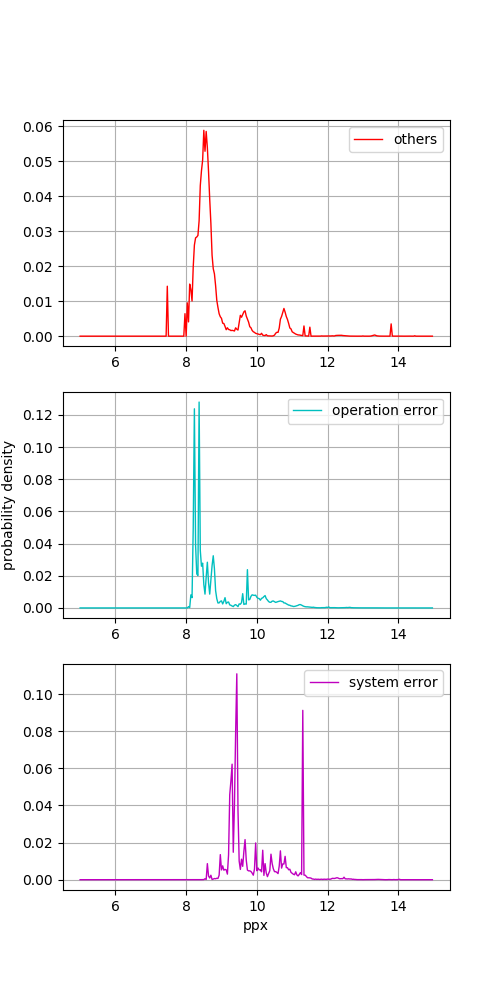}
		\label{fig:scrmweb}
		}
	\caption{The distribution of the perplexity of log entries drawn from different log types.}
	\label{fig:scrm}
	\end{figure}

	Our empirical study on the perplexity suggests that the perplexity score alone is a strong indicator of the natural of the log entry. Figure \ref{fig:scrmapp} and Figure \ref{fig:scrmweb} illustrate the distributions of the perplexity that drawn from the  system log carrying different labels. 
	
	Both figures show that there is exist a peak in the distribution in the perplexity range from 8 to 9, regardless of the pre-set label of the message type. Moreover, log messages with the label {\em system error} tend to have another large spike in the high-perplexity range (around 11.5). This spike is more obvious in Figure \ref{fig:scrmapp}, which is much higher than what it has around 8 to 9. We assume that a special and rarely-seen type of system error tend to have a very long trace of exceptions into the calling stack, causing both its lexicon and word sequence more obscure for the language model. 
	The threshold to cut the high-frequency message types is a hyper-parameters, which balance the trade-off between the recall and speed acceleration. In practice we choose a threshold of $11$ that filter out all common logging messages for further analysis. 
	
\section{Log Classification}

	The second stage of our method consists of two steps. We first use a classic topic modeling technique to leverage the unlabeled data to construct a semantic space, then apply a multilayer perceptron neural network to perform a supervised learning for the three-way classification task. 

	\subsection{LDA Topic Modeling}
	
\begin{figure}[h]
	\centering
	\includegraphics[width=0.55\textwidth]{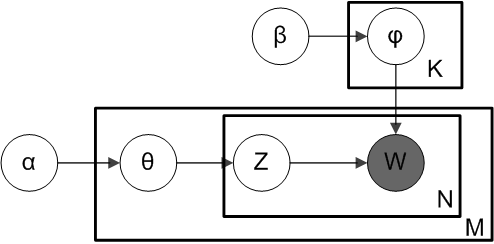}
	\caption{The plate notation for LDA with Dirichlet-distributed topic-word distributions.}
	\label{fig:lda}
	\end{figure}

	The language model based filter outputs a filtered set of log entries, that are still in the raw text form, which requires further transformation to make it suitable for classification tasks. 
	Latent Dirichlet allocation (LDA) \cite{blei2003latent} is a directed graphic model, that is based on a Bag-of-Word representation of texts, makes the assumptions that each of the word occurrence is a sample drawn from a multi-nominal distribution of the vocabulary, conditioned on the topic the text wants to express, and uses the Dirichlet as its conjugate prior for smoothing. In practice, we often use a extended version of LDA which has an extra Dirichlet-distributed topic-word distribution, as shown in Figure \ref{fig:lda}. 
	 
	The three values of shown in Figure \ref{fig:lda} are $K$ the number of topics, $M$ the number of documents, and $N$ the number of words in each document. 
	Besides, the interpretation of several key parameters are as follows:
	\begin{itemize}
		\item $\boldsymbol{\alpha}$ is the Dirichlet prior on the per-document topic distributions, 
		\item $\boldsymbol{\beta}$ is the Dirichlet prior on the per-topic word distribution, 
		\item $\theta_m$ is the topic distribution for document $m$, 
		\item $\phi_k$ is the word distribution for topic $k$, 
		\item $z_{m,n}$ is the topic for the $n$-th word in document $m$, 
 	    \item and ${\displaystyle w_{m,n}}$ is the specific word. 
	\end{itemize}
	The only observable variables (i.e., the node with gray background) is the occurrence of $n$ in document $m$, denoted as $w_{m,n}$. 
	We use the gensim\footnote{https://radimrehurek.com/gensim/index.html} implementation of the LDA in our experiment \cite{rehurek_lrec}, and choose hyper-parameter $K$ (i.e., the number of topics) to be $100$, $\alpha_{k=1\dots K} = \frac{1.0}{K}$, and $\beta_{w=1\dots V}=\frac{1.0}{K}$. 
	
	\subsection{Classification by MLP}

	Multilayer Perceptron (MLP) is one of the basic type of feed-forward neural network architecture with densely connected layers one after another, whose weights are tuned by a supervised learning technique called back-propagation \cite{rumelhart1985learning}.  
	The input to the MLP's input units are the $K$ dimensional semantic vectors derived from the transoformation of the Bag-of-Word representation of the original text by the previously learnt LDA topic model.
	In our experiment, we create a 5 layers of the densely connected feedforward network, consisting of 3 hidden units in each layer, with ReLU as the activation function, and the $softmax$ for the last layer. 

\section{Empirical Results}

	We select two consecutive days' log data from two representative system modules, namely {\em scrmweb} and {\em scrmapp} of a national-wide telecom company. The volume sizes of these log data files are about 10GB and 200GB, containing 8M+ and 90M+ message instances respectively. To ensure that our analyzer is able to process the day-round logging output sufficiently fast, the average processing time of one day's log data must be shorter than 24 hours. 
	In our method, the major time consumption lies in the first stage, which uses the language model to pre-filter out most of the logging data. In the experiment, we allocate 20/30 cores for language modeling, Multi-cores of E5 2630v4 and 1 Nvidia Titan XP for classification. 
	Table \ref{tbl:time} shows the log processing time consumption of the two modules. 
	
	
	\begin{table}[h]
	\centering
	\caption{The log processing time consumption of the $ppx$-based pre-filtering. }
	\label{tbl:time}
    	\begin{tabular}{l||c|c|c|c|c}
    		\hline
    		Sub-module	& Volume	& \#Core Alloc	& Avg CPU 	& Total time 	& Avg Speed \\
    		\hline
    		scrmapp		& $\sim$200GB	& 20		& 1403\%	& $\sim$23:59	& $\sim$165KB/s per-core\\
    		scrmweb		& $\sim$10GB		& 30 		& 1923\%	& $\sim$1:09	& $\sim$124KB/s per-core\\
    		\hline
    	\end{tabular}
	\end{table}
	
	Our experiment shows that even by the simple 3-gram language model, the pre-filtering stage still takes most of the total time consumption. For a two days' data, the pre-filtering time consumption is almost an entire day. The average processing speed, depending on the different module it works on, varies from 124KB/s per-core to 165KB/s per-core, but still several orders of magnitude faster than the inference stage of the LDA + MLP classifier.  	

\begin{figure}[h]
	\centering
	\includegraphics[width=0.70\textwidth]{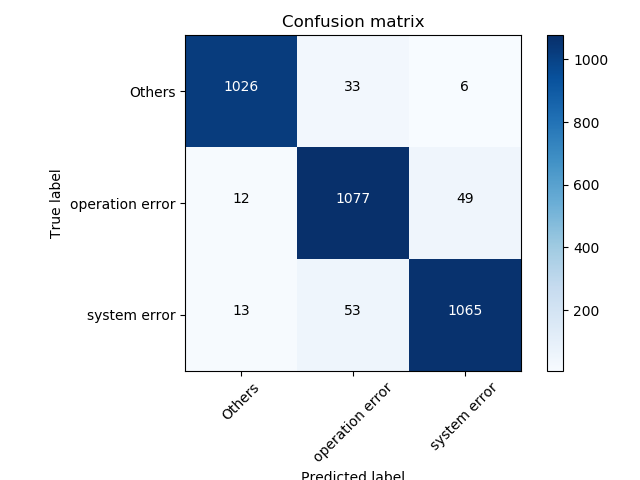}
	\caption{The confusion matrix of the 3-way classification result on our test dataset. }
	\label{fig:confusion}
	\end{figure}

	Our second experiment considers the classification accuracy of our method. We conducts our experiment on the `scrmweb` sub-module. 
	Table \ref{tbl:label-dist} tells the distribution of the pre-set labels of the data. 
	\begin{table}[h]
	\centering
	\caption{The distribution of the pre-set labels of the data in sub-module `scrmweb`.}
	\label{tbl:label-dist}
		\begin{tabular}{l||c|c}
		\hline
		Log Type	& \#Instance		& Percentage \\
		\hline
		Normal		& 5.6M+		& 70.8\% \\
		Operation	& 2.2M+		& 27.8\% \\
		System		& 0.1M+		& 1.4\% \\
		\hline
		\end{tabular}
	\end{table}
	To create a ground-truth dataset as our test bed, we select a subset of the two-day data, consisting of 20,000 instances (about 25M in size). We allocates 5/6 of the instances for training, and the left 1/6 for testing. We create {\em True labels} for these instances, by post annotation with manual efforts, which may differ from the pre-set label by the target system. 
    Figure \ref{fig:confusion} is a confusion matrix of our final classification performance (3,334 instances). Rows in the table are the numbers of the instances for each true labeled class, columns in the table are the numbers of the instances for each predicated class of our method. Our empirical experiment shows that the average Precision/Recall/F-score reach 0.973, 0.976 and 0.975 respectively. 

	Notice that most of the log data are pre-filtered, only a small portion of the entire dataset is left for the LDA topic modeling and MLP classifier. The processing speed in this stage is $\sim$1.9KB/s per-core, which is approximately 100 times slower than the language model based filtering stage, making pre-filtering stage an essential part of the entire process.

\section{Conclusions}

	Designing a logging analyzer for a large scale online system faces several key challenges, including the vague semantics of the logging message to be recovered, the need of a sufficiently fast processing speed, and above all, the lack of prior knowledge of how the developers of the target system choose the templates and vocabularies. Based on these considerations, we propose a ``least prior knowledge'' principle for designing such a system, and present an implementation based on several key natural language processing techniques. We use a fast n-gram language model to pre-filter the trivial types of messages, which considers both its use of template and lexicon. Then we use Latent Dirichlet allocation (LDA) topic modeling method to construct a semantic space from a large repository of unlabeled (ground truth labeling, rather than the pre-set misleading labels give by the target system) bag-of-word representation of the messages. Finally, a standard feed-forward multilayer-perceptron is used to perform the three-way classification, with the minimum amount of human supervision. 

\bibliographystyle{splncs}
\bibliography{main}

\begin{thebibliography}{10}

\bibitem{silverstein1999analysis}
Silverstein, C., Marais, H., Henzinger, M., Moricz, M.:
\newblock Analysis of a very large web search engine query log.
\newblock In: ACm SIGIR Forum. Volume~33., ACM (1999)  6--12

\bibitem{xu2009detecting}
Xu, W., Huang, L., Fox, A., Patterson, D., Jordan, M.I.:
\newblock Detecting large-scale system problems by mining console logs.
\newblock In: Proceedings of the ACM SIGOPS 22nd symposium on Operating systems
  principles, ACM (2009)  117--132

\bibitem{yen2013beehive}
Yen, T.F., Oprea, A., Onarlioglu, K., Leetham, T., Robertson, W., Juels, A.,
  Kirda, E.:
\newblock Beehive: Large-scale log analysis for detecting suspicious activity
  in enterprise networks.
\newblock In: Proceedings of the 29th Annual Computer Security Applications
  Conference, ACM (2013)  199--208

\bibitem{kobayashi2014towards}
Kobayashi, S., Fukuda, K., Esaki, H.:
\newblock Towards an nlp-based log template generation algorithm for system log
  analysis.
\newblock In: Proceedings of The Ninth International Conference on Future
  Internet Technologies, ACM (2014) ~11

\bibitem{charniak1996statistical}
Charniak, E.:
\newblock Statistical language learning.
\newblock MIT press (1996)

\bibitem{ponte1998language}
Ponte, J.M., Croft, W.B.:
\newblock A language modeling approach to information retrieval.
\newblock In: Proceedings of the 21st annual international ACM SIGIR conference
  on Research and development in information retrieval, ACM (1998)  275--281

\bibitem{salvetti2006weblog}
Salvetti, F., Nicolov, N.:
\newblock Weblog classification for fast splog filtering: A url language model
  segmentation approach.
\newblock In: Proceedings of the Human Language Technology Conference of the
  NAACL, Companion Volume: Short Papers, Association for Computational
  Linguistics (2006)  137--140

\bibitem{du2017deeplog}
Du, M., Li, F., Zheng, G., Srikumar, V.:
\newblock Deeplog: Anomaly detection and diagnosis from system logs through
  deep learning.
\newblock In: Proceedings of the 2017 ACM SIGSAC Conference on Computer and
  Communications Security, ACM (2017)  1285--1298

\bibitem{blei2003latent}
Blei, D.M., Ng, A.Y., Jordan, M.I.:
\newblock Latent dirichlet allocation.
\newblock Journal of Machine Learning Research \textbf{3} (2003)  993--1022

\bibitem{rehurek_lrec}
{\v R}eh{\r u}{\v r}ek, R., Sojka, P.:
\newblock {Software Framework for Topic Modelling with Large Corpora}.
\newblock In: {Proceedings of the LREC 2010 Workshop on New Challenges for NLP
  Frameworks}, Valletta, Malta, ELRA (May 2010)  45--50

\bibitem{rumelhart1985learning}
Rumelhart, D.E., Hinton, G.E., Williams, R.J.:
\newblock Learning internal representations by error propagation.
\newblock Technical report, California Univ San Diego La Jolla Inst for
  Cognitive Science (1985)

\end{thebibliography}

\end{document}